\documentclass[smallcondensed,natbib]{svjour3}
\smartqed  
\usepackage{graphicx}
\usepackage{mathptmx}      
\usepackage{amsfonts}
\usepackage{amsmath}
\usepackage{amssymb}
\usepackage[breaklinks=true]{hyperref}
\begin{document}

\title{Tracking azimuthons in nonlocal nonlinear media}


\author{F.~Maucher \and
        D.~Buccoliero \and
        S.~Skupin \and
        M.~Grech \and
        A.~S.~Desyatnikov \and
        W.~Krolikowski
}


\institute{F.~Maucher \and S.~Skupin \and M.~Grech \at
           Max Planck Institute for the Physics of Complex Systems, N\"othnitzer Str.~38,
           01187 Dresden, Germany
           \and
           D.~Buccoliero \and A.~S.~Desyatnikov \at
           Nonlinear Physics Center, Research School of Physics and Engineering,
           Australian National University, Canberra, ACT 0200, Australia
           \and
           S.~Skupin \at
           Institute of Condensed Matter Theory and Solid State Optics, Friedrich-Schiller-University,
           Max-Wien-Platz 1, 07743 Jena, Germany
           \and
           W.~Krolikowski \at
           Laser Physics Center, Research School of Physics and Engineering,
           Australian National University, Canberra, ACT 0200, Australia
}


\maketitle

\begin{abstract}
We study the formation  of azimuthons, i.e., rotating spatial solitons, in
media with nonlocal focusing nonlinearity. We show that whole families
of these solutions can be found by considering internal modes of classical non-rotating stationary solutions, namely
vortex solitons. This offers an exhaustive method to identify azimuthons in a given nonlocal medium.
We demonstrate formation of
azimuthons  of different vorticities and explain their properties by considering the
strongly nonlocal limit of accessible solitons.

\end{abstract}

\section{Introduction}
\label{intro}
There has been growing interest in studies of propagation of optical beams  in nonlocal media. These are media
where the nonlinear response of the material in a specific spatial location is determined not only by the wave
intensity in the same location but also in its neighborhood. The extent of this neighborhood in comparison to the beam
width determines the degree of nonlocality. The nonlinear nonlocal response  appears to be ubiquitous to many physical settings.
For instance, it is common to media where certain transport processes such as
heat~\citep{Dabby:apl:13:284,Litvak:sjpp:1:31,Davydova:ujp:40:487} or charge transfer~\citep{Calvo:epl:60:847},
diffusion and/or drift of
atoms~\citep{Suter:pra:48:4583,Skupin:prl:98:263902}  are responsible for the nonlinearity. It also occurs in
systems involving long-range interaction of atoms or molecules as it is the case of nematic liquid
crystals~\citep{Conti:prl:92:113902,Conti:prl:91:073901} or dipolar Bose-Einstein
condensate~\citep{Goral:pra:61:051601,Nath:pra:76:013606,Koch:np:4:218}. It has been shown that nonlocal
nonlinear response has profound
consequences on the wave propagation and formation of localized structures~\citep{Krolikowski:jobsqo:6:288}. In
particular, nonlocality prevents collapse by providing a stabilization mechanism and enables  robust existence  of various
types of localized structures and spatial
solitons~\citep{Kolchugina:jetpl:31:304,Bang:pre:66:046619,Briedis:oe:13:435,Skupin:pre:73:066603,
Lashkin:pra:75:043607, Lashkin:pla:366:422}. In
local nonlinear media the wave   perturbation in a particular place  affects the nonlinearity which in turn, influences
the wave itself often instigating its breakup or spatial transformation \citep{Desyatnikov:po:47:291}. 
On the other hand,  in  nonlocal media such
perturbation is spatially averaged, and hence has a much weaker impact 
 on  the wave itself, leading to its
stabilization. In particular, it has been shown that  nonlocality
 support stable propagation of optical vortices, and multi-peak solitonic structures  which are structurally unstable in
material with local response \citep{Buccoliero:ol:33:198,Buccoliero:pb:394:351,Buccoliero:prl:98:053901}.
A range of particular types 
of fundamental as well as higher order  nonlocal
solitons and their interactions  have been even demonstrated 
experimentally in materials with nonlocal response of
thermal origin~\citep{Rotschild:ol:31:3312,Rotschild:np:2:769}. Recently, it has been also shown
theoretically  that
spatial nonlocal response enables realization of the 
so called azimuthons i.e. multiple peak ring-shaped solitons which exhibit
angular rotation in propagation~\citep{Desyatnikov:prl:95:203904,
Lopez:ol:31:1100,Lopez:oe:14:7903,Skupin:oe:16:9118}.
While few specific types of azimuthons have been
investigated in various nonlocal models, using variational techniques
mentioned above as well as numerical relaxation procedure \citep{Lashkin:pra:77:025602,Lashkin:pra:78:033603}, no
general approach to find stable nonlocal azimuthons has
been demonstrated so far.
In this  work we 
study the formation of azimuthons in nonlocal media
with gaussian response. We show that whole families of these
 solitons can be tracked down by analyzing bifurcations
originating from the nonlinear optical potential of vortex solitons.

\section{Azimuthons \label{rotsol}}

We consider physical systems governed by the two-dimensional 
nonlocal nonlinear Schr{\"o}\-dinger equation
\begin{equation}
\label{NLS}
i \frac{\partial}{\partial z} \psi +  \Delta_{\perp}\psi  + \theta \psi = 0.
\end{equation}
where $\theta$ represents the spatially nonlocal nonlinear response of the medium. Its form depends on the details of a particular physical system.
In the following, we will assume that the nonlinear response $\theta$ can be
expressed in terms of the nonlocal response function $R(r)$
\begin{equation}
\label{NL_R_int}
\theta = \iint R(|\vec{r}-\vec{r}^{\prime}|) \left|\psi(\vec{r}^{\prime},z)\right|^2d^2\vec{r}^{\prime},
\end{equation}
where $\vec{r}=x\vec{e}_x+y\vec{e}_y$ denotes the transverse coordinates.
In this work we will use the so-called
Gaussian model of nonlocality as an illustrative example,
\begin{equation}
\label{NL_gaussian_int}
\theta = \frac{1}{2\pi}\iint\mathrm{e}^{-\frac{|\vec{r}-\vec{r}^{\prime}|^2}{2}}
\left|\psi(\vec{r}^{\prime},z)\right|^2d^2\vec{r}^{\prime}.
\end{equation}
However, the proposed solutions should exist in many other nonlocal models
\citep{Litvak:sjpp:1:31,Rotschild:np:2:769,Suter:pra:48:4583,Skupin:prl:98:263902,Assanto:ieeejqe:39:13,
Conti:prl:92:113902,Peccianti:np:2:737,Denschlag:sc:287:97,Pedri:prl:95:200404,Koch:np:4:218}.

Azimuthons are a straightforward generalization of the usual ansatz for
stationary solutions (solitons) \citep{Desyatnikov:prl:95:203904}.
They represent spatially rotating structures and hence involve an
additional parameter, the angular frequency $\Omega$ (see also \cite{Skryabin:pre:66:055602})
\begin{equation}
\label{azimuthon_ansatz}
\psi (r,\phi ,z) = U(r,\phi -\Omega z)\mathrm{e}^{i\lambda z},
\end{equation}
where $U$ is the complex amplitude function and $\lambda$ the propagation constant.
For $\Omega=0$, azimuthons become ordinary (nonrotating) solitons.
The  simplest example of a  family of azimuthons is the one connecting the dipole soliton
with the single charged vortex soliton \citep{Lopez:ol:31:1100}. A single charged vortex
consists of two equal-amplitude
dipole-shaped structures with the relative phase of $\pi/2$ 
representing real and imaginary part of $U$.
If these two components differ in amplitudes the resulting structure
 forms a ''rotating dipole'' azimuthon.  If one of
the components  is zero we deal with the nonrotating dipole soliton.
In the following we will denote the  amplitude ratio of these two  vortex components  by $\alpha$, which
also determines the angular modulation depth  of the resulting ring-like structure
by ``$1-\alpha$''. When higher order  (e.g. single charged triple-hump) azimuthons are concerned, we can not always
identify the angular modulation depth with amplitude ratios of real and imaginary part of $U$. Hence we define the
generalized structural parameter $\alpha$ as
\begin{equation}
\label{define_alpha}
\alpha=1-\frac{\min_{\phi}|U(r_{\rm max},\phi)|}{|U(r_{\rm max},\phi_{\rm max})|},
\end{equation}
where the tuple $(r_{\rm max},\phi_{\rm max})$ denotes the coordinates of the maximum value $\max_{r,\phi}{|U|}$.

After inserting the  ansatz~(\ref{azimuthon_ansatz}) into Eqs.~(\ref{NLS}) and
(\ref{NL_R_int}), multiplying  with $U^*$ and $\partial_{\phi}U^*$ resp., and integrating
over the transverse coordinates we end up with
\begin{subequations}
\label{int_system}
\begin{align}
-\lambda M + \Omega L_z +I+N & =0 \label{int_system_a}\\
- \lambda L_z + \Omega M^{\prime} + I^{\prime} + N^{\prime} & =0.
\end{align}
This system relates the propagation constant $\lambda$ and the rotation
frequency $\Omega$ of the azimuthons to integrals over their stationary
amplitude profiles, namely
\begin{align}
M & = \iint \left|U(\vec{r})\right|^2d^2\vec{r} \\
L_z & = -i \iint U^*(\vec{r})
\frac{\partial}{\partial\phi}U(\vec{r})d^2\vec{r} \\
I & = \iint U^*(\vec{r})
\Delta_{\perp} U(\vec{r})d^2\vec{r} \\
N & = \iiiint R(|\vec{r}-\vec{r}^{\prime}|)
\left|U(\vec{r}^{\prime})\right|^2\left|U(\vec{r})\right|^2d^2\vec{r}^{\prime}d^2\vec{r}
\\
M^{\prime} & = \iint
\left|\frac{\partial}{\partial\phi}U(\vec{r})\right|^2d^2\vec{r} \\
I^{\prime} & = i \iint \left[ \frac{\partial}{\partial\phi} U^*(\vec{r}) \right]
\Delta_{\perp} U(\vec{r})d^2\vec{r} \\
N^{\prime} & = i \iiiint R(|\vec{r}-\vec{r}^{\prime}|)
\left|U(\vec{r}^{\prime})\right|^2 \left[ \frac{\partial}{\partial\phi}
U^*(\vec{r}) \right] U(\vec{r})d^2\vec{r}^{\prime}d^2\vec{r}.
\end{align}
\end{subequations}
The first two quantities have straightforward physical meanings, namely ''mass'' ($M$) and ''angular
momentum'' ($L_z$). We can formally solve for the rotation frequency and obtain (for an alternative
derivation see \citep{Rozanov:os:96:405}
\begin{equation}
\label{omega_exact}
\Omega=\frac{M\left(I^{\prime}+N^{\prime}\right)-L_z\left(I+N\right)}{L_z^2-MM^{\prime}}.
\end{equation}
Note that this expression is undetermined for a vortex beam.
For
$\alpha=1$ [vortex soliton $V(r)\exp(iq\phi+i\lambda_0 z$)], we can assume any value for $\Omega$ by
just shifting the propagation constant $\lambda=\lambda_0+\Omega$ accordingly
($\lambda_0$ accounts for the propagation
constant in the non-rotating laboratory
frame). However, with
respect to a particular azimuthon in the limit $\alpha \rightarrow 1$, 
the value of $\Omega$ is fixed.
In what follows, we denote this value by $\Omega|_{\alpha=1}$.

\section{Internal modes and azimuthons \label{intmodaz}}

In this section we will discuss  the formation
of azimuthons via the  process of bifurcation from a stationary non-rotating
soliton solution, namely a vortex. We assume a certain deformation of the soliton profile while
going over from the vortex to azimuthons in the limit  $\alpha \rightarrow 1$. Therefore it has
to be the shape of vortex deformation which determines $\Omega$, since a
vortex formally allows for all possible rotation frequencies (see the discussion on shifting $\lambda$
at the end of Sec.~\ref{rotsol}).

Let us now look at the  azimuthon originating (bifurcating) from a vortex soliton with charge $q$.
 For this purpose, we recall the eigenvalue problem
for internal modes of the  nonlinear potential $\theta$ which is usually treated in the context of
 linear stability of nonlinear soliton solutions \citep{Firth:prl:79:2450,Desyatnikov:po:47:291}. We
introduce a small
 perturbation $\delta V$ to the vortex soliton $V$, 
\begin{equation}
\psi = \left(V+\delta V\right)\mathrm{e}^{iq\phi+i\lambda_0z},
\end{equation}
plug it into Eqs.~(\ref{NLS}) and
(\ref{NL_R_int}) and  linearize those equations with respect to the perturbation.
Note that the perturbation $\delta V(r,\phi,z)$ is
complex, whereas the vortex profile $V(r)$ is real (w.l.o.g.).
The resulting evolution equation for the perturbation $\delta V$ is then given by
\begin{equation}
\begin{split}
\left[ i \frac{\partial}{\partial z} -\lambda_0 +
\frac{1}{r}\frac{\partial}{\partial r}\left(r\frac{\partial}{\partial
r}\right) + \frac{1}{r^2} \left(\frac{\partial}{\partial \phi} +iq
\right)^2 + \iint R(|\vec{r}-\vec{r}^{\prime}|) V^2(r^{\prime})d^2\vec{r}^{\prime} \right]\delta V & \\
+ V\iint R(|\vec{r}-\vec{r}^{\prime}|) V(r^{\prime}) \left[ \delta V(\vec{r}^{\prime},z) + \delta
V^*(\vec{r}^{\prime},z) \right] d^2\vec{r}^{\prime} &=0.
\end{split}
\end{equation}
With the ansatz
\begin{equation}
\delta V = \delta V_1(r)\mathrm{e}^{im\phi+i\kappa z} + \delta V^*_2(r)\mathrm{e}^{-im\phi-i\kappa^* z}
\end{equation}
we derive the eigenvalue problem for the internal modes
\begin{subequations}
\label{Eigen_vortex}
\begin{align}
 \left[ \frac{1}{r}\frac{\partial}{\partial r}\left(r\frac{\partial}{\partial r}\right) - \frac{(m+q)^2}{r^2}
 -\lambda_0 + \iint R(|\vec{r}-\vec{r}^{\prime}|) V^2(r^{\prime})d^2\vec{r}^{\prime} \right] \delta V_1 & \nonumber \\ 
 +V\iint R(|\vec{r}-\vec{r}^{\prime}|)V(r^{\prime}) \left[\delta V_1(r^{\prime}) + \delta V_2(r^{\prime}) \right]
 \cos[m(\phi-\phi^{\prime})] d^2\vec{r}^{\prime} & = \kappa \delta V_1 \\
 -\left[\frac{1}{r}\frac{\partial}{\partial r}\left(r\frac{\partial}{\partial r}\right) - \frac{(m-q)^2}{r^2}
 -\lambda_0 + \iint R(|\vec{r}-\vec{r}^{\prime}|) V^2(r^{\prime})d^2\vec{r}^{\prime} \right] \delta V_2 & \nonumber \\ 
 -V\iint R(|\vec{r}-\vec{r}^{\prime}|)V(r^{\prime}) \left[\delta V_2(r^{\prime}) +\delta V_1(r^{\prime}) \right]
\cos[m(\phi-\phi^{\prime})] d^2\vec{r}^{\prime} & = \kappa \delta V_2.
\end{align}
\end{subequations}
Note that since
$|\vec{r}-\vec{r}^{\prime}|=\sqrt{r^2+r^{\prime2}
-2rr^{\prime}\cos(\phi-\phi^{\prime})}$,
all integrals in (\ref{Eigen_vortex}) are independent of $\phi$.
Real-valued eigenvalues of Eq.~(\ref{Eigen_vortex}) ($\kappa=\kappa^*$) are termed
orbitally stable and the corresponding eigenvector $(\delta V_1, \delta V_2)$
can be chosen as real.
If we perturb the vortex $V$ with an orbitally stable
eigenvector, the resulting wave-function $\psi$ can be written in the form of
Eq.~(\ref{azimuthon_ansatz}) with $\Omega=-\kappa/m$ and $\lambda=\lambda_0-q\kappa/m$.
Thus, it is possible to
construct azimuthons in the vicinity of the vortex ($\alpha \approx 1$)
from $\delta V$:
\begin{equation}
\label{azimvor}
U (r,\phi)|_{z=0} = \left[ V(r) + A_r \delta V_1(r)\mathrm{e}^{im\phi} + A_r\delta
V_2(r)\mathrm{e}^{-im\phi}\right]\mathrm{e}^{iq\phi}.
\end{equation}
Used as an initial condition  in the propagation equation (\ref{NLS}) this object is expected to rotate with an angular frequency
$\Omega|_{\alpha=1}=-\kappa/m$.
Here, $A_r>0$ was introduced as the amplitude of the perturbation $\delta V$ with respect to $V$.
Since we are operating in a linearized system, the amplitude of the perturbation as a solution of
Eq.~(\ref{Eigen_vortex}) is not fixed (just the
ratio between the components $\delta V_1$ and $\delta V_2$ is prescribed), but will eventually determine the value of
the structural parameter $\alpha$. Generally speaking, the smaller the resulting $\alpha$ the greater the error in the
constructed initial condition. However, the great robustness of the azimuthons, at least in the Gaussian model, allows
one to use the initial
condition~(\ref{azimvor}) for quite large perturbation amplitudes $A_r$. Those strongly perturbed initial
conditions result in  oscillations of the azimuthon upon propagation. However, the azimuthon is structurally stable and does not
decay into other soliton solutions like the single-hump ground state. Moreover, such initial conditions play a role of
excellent ''initial guesses'' for solver routines to find numerically exact azimuthons.

\section{Higher order azimuthons}

In a recent  publication we have used the approach 
presented above to characterize the rotating dipole azimuthon,
which connects the single charged vortex ($q=1$) to the stationary dipole soliton \citep{Skupin:oe:16:9118}. As already
mentioned there, solving the eigenvalue problem~(\ref{Eigen_vortex}) can be used as an exhaustive method for finding
families of azimuthons which originate from a vortex soliton. However, it should be stressed that  not all orbitally
stable eigenvalues can be linked to a family of azimuthons. This is obvious for eigenvalues with
$|\kappa|>|\lambda_0|$ in the continuous part of the spectrum. Hence, we can
conclude that $|\Omega|_{\alpha=1}|<|\lambda_0|/m$ for azimuthons in the vicinity of the vortex ($\alpha\simeq1$).
Note that the parameter $m$ determines  the number of humps of the rotating structure.

Looking  at a higher order azimuthon, e.g., a single charged rotating triple hump ($q=1,m=3$),
the natural question one may pose is
whether this particular family of azimuthons is connected to a second
bifurcation from a vortex soliton, as predicted by variational calculations
\citep{Desyatnikov:prl:95:203904,Lopez:oe:14:7903}.
In the case of the rotating
double hump we had a natural candidate, namely the stationary dipole; 
the existence of an analogous solution like a
stationary tripole is not evident. It turns out that the rotating triple hump azimuthon with lowest absolute rotation frequency $\Omega$ connects single
and double charged vortex with opposite sign of charge, in contrast to variational
predictions \citep{Lopez:oe:14:7903}.
At least in the highly nonlocal regime the rotating frequency does not change much when we
follow the family with constant mass (here $M=630$ and $-3>\Omega>-3.5$), we do not find a solution with $\Omega=0$.
What we do find is a solution with vanishing angular momentum $L_z$, because the two limiting vortices have opposite
sign of charge.

\begin{figure}
\centerline{\includegraphics[width=.3\columnwidth]{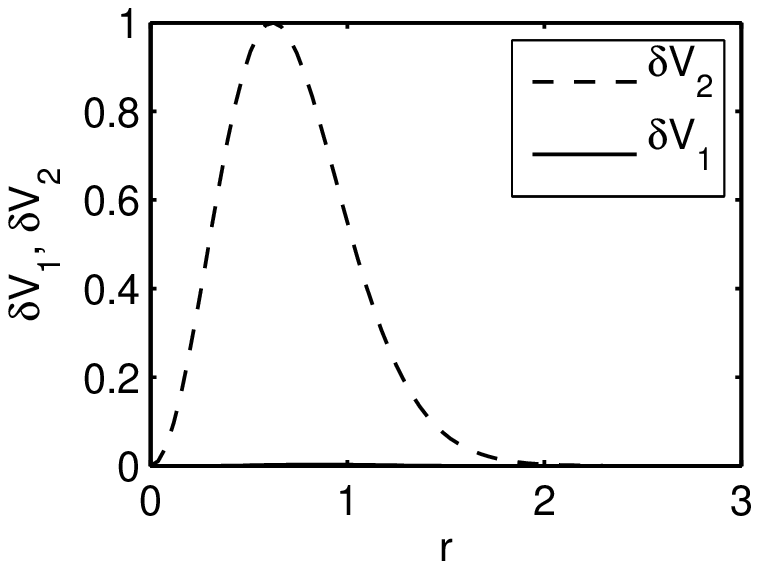}
\hspace{0.25cm}\includegraphics[width=.3\columnwidth]{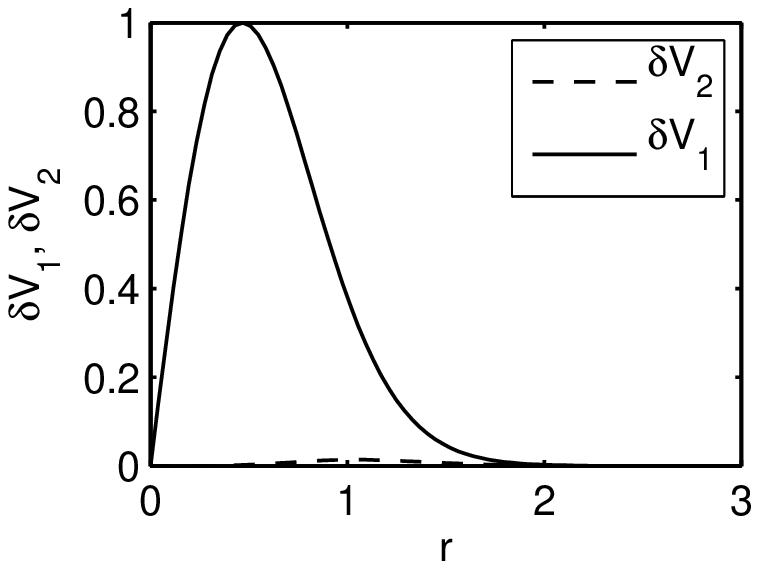}}
\caption{\label{fig.internalmodes3} Internal modes ($m=3$) of the single charges vortex ($q=1$) with eigenvalue
$\kappa=10.3$ (left) and the double charged vortex ($q=-2$) with eigenvalue $\kappa=9.1$ (right). $\delta V_{2}$ clearly dominates the nearly 
invisible $\delta V_{1}$ component in the left picture, whereas the opposite is the case in the right one.}
\end{figure}

\begin{figure}
\centerline{\includegraphics[width=.3\columnwidth]{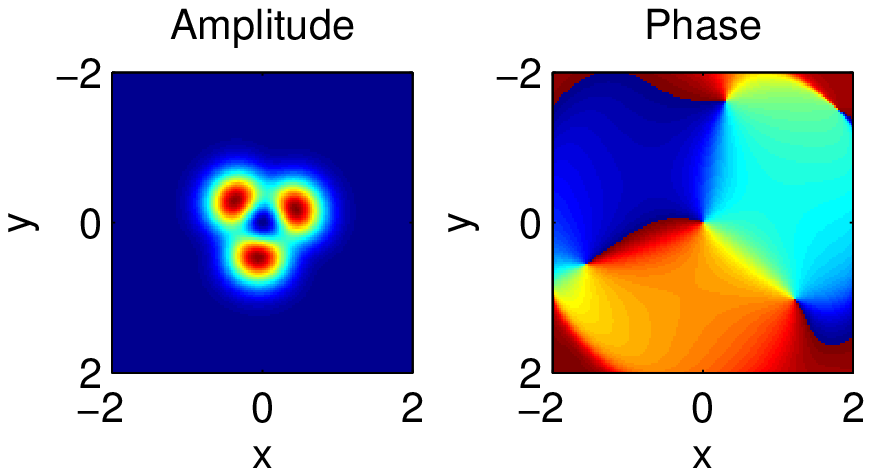}
\hspace{0.25cm}\includegraphics[width=.3\columnwidth]{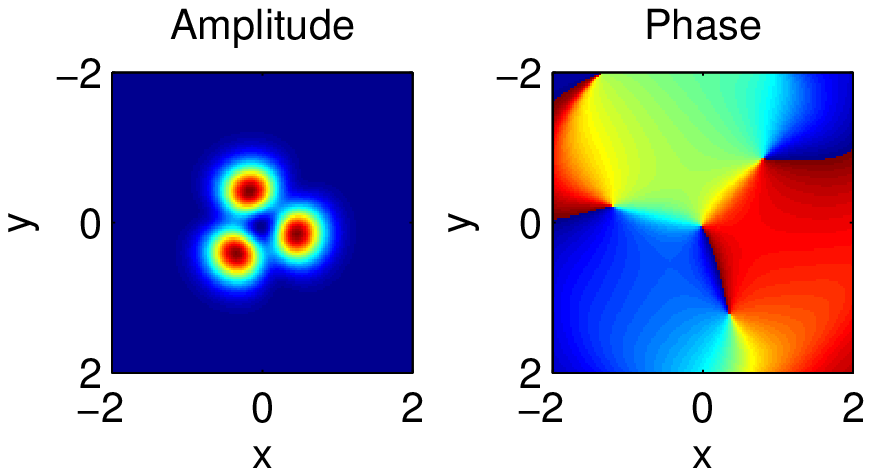}
\hspace{0.25cm}\includegraphics[width=.3\columnwidth]{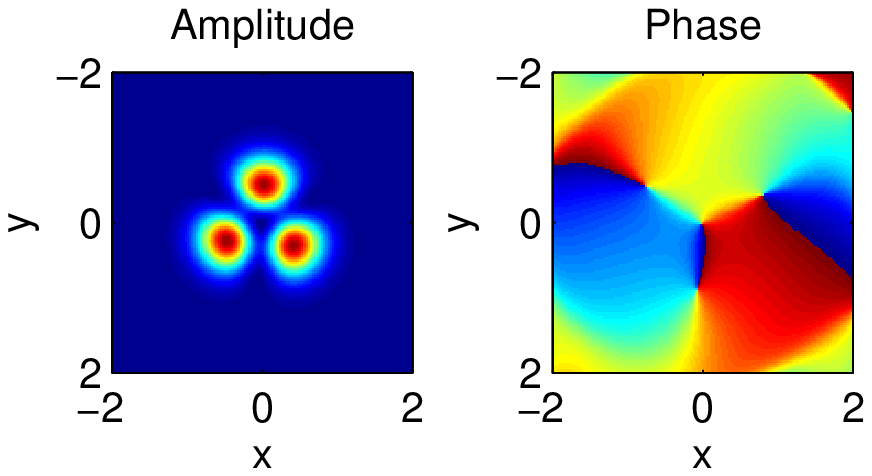}}
\vspace{0.25cm}
\centerline{\includegraphics[width=.3\columnwidth]{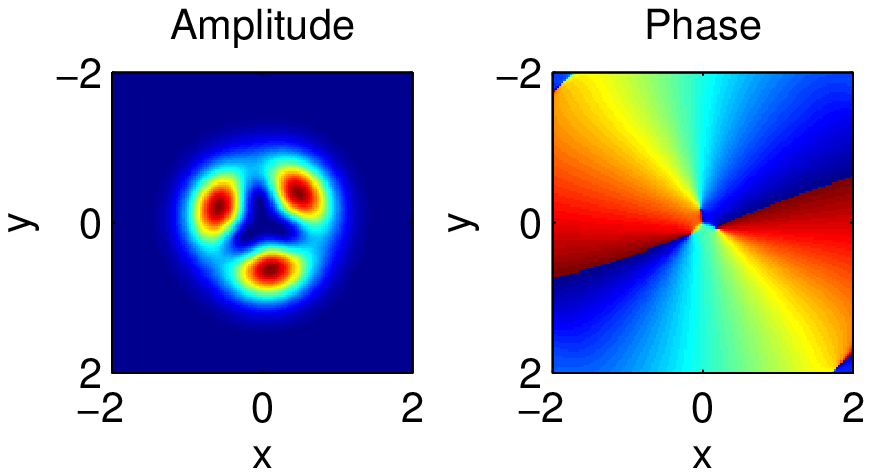}
\hspace{0.25cm}\includegraphics[width=.3\columnwidth]{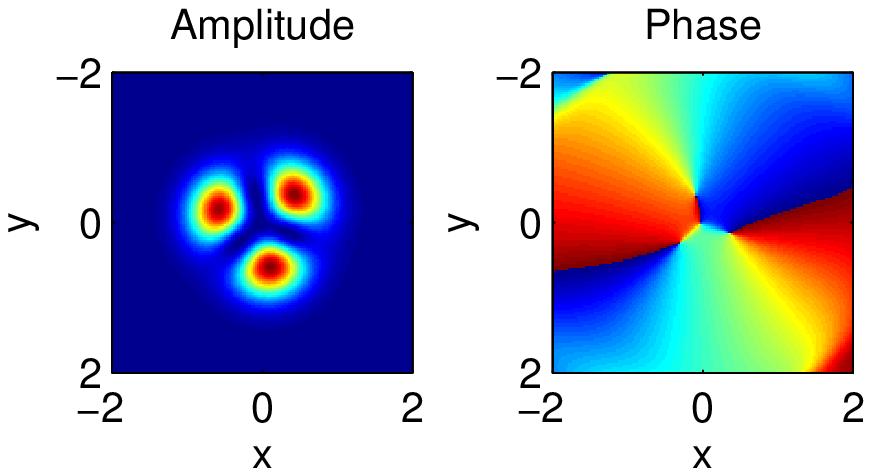}
\hspace{0.25cm}\includegraphics[width=.3\columnwidth]{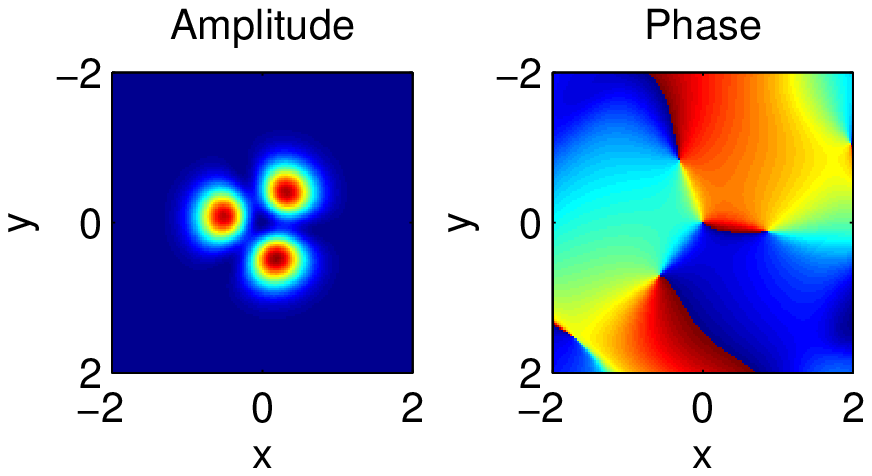}}
\caption{\label{fig.triple} (First row) Tracking the family starting from the single charged vortex ($q=1$), the
respective angular momenta and structural parameter are (from left to right) $L_z=450$, $\alpha=0.65$; $L_z=270$,
$\alpha=0.53$; and $L_z=0$, $\alpha=0.4$.
(Second row) Tracking the family starting from the double charged
vortex ($q=-2$); $L_z=-1100$, $\alpha=0.55$; $L_z=-770$, $\alpha=0.3$; and
$L_z=0$, $\alpha=0.4$ (from left to right).}
\end{figure}

Analysis of internal modes of both vortices ($V$) (charge $q=1$ and $q=-2$, charge of perturbation $m=3$)
reveals eigenvalues and eigenvectors where the family of azimuthons emerges. Figure \ref{fig.internalmodes3} shows the
two eigenvectors for mass $M=630$, the resulting rotation frequency is $\Omega=-\kappa/3$.
Now we can track the family starting from both vortices ($q=1$ and $q=-2$) perturbed with the appropriate eigenvectors
and constant mass $M=630$. Equation~(\ref{azimvor}) serves as initial condition (we increase the perturbation
amplitude) to a Newton solver. We follow both branches till we reach $L_z=0$, where the two solutions
coincide (see last plot in each row of Fig.~\ref{fig.triple}). Interestingly, we observe that the triple hump azimuthon
with $L_z=0$
is not the one with maximum modulation depth $1-\alpha$.

\begin{figure}
\centerline{\includegraphics[width=.2\columnwidth]{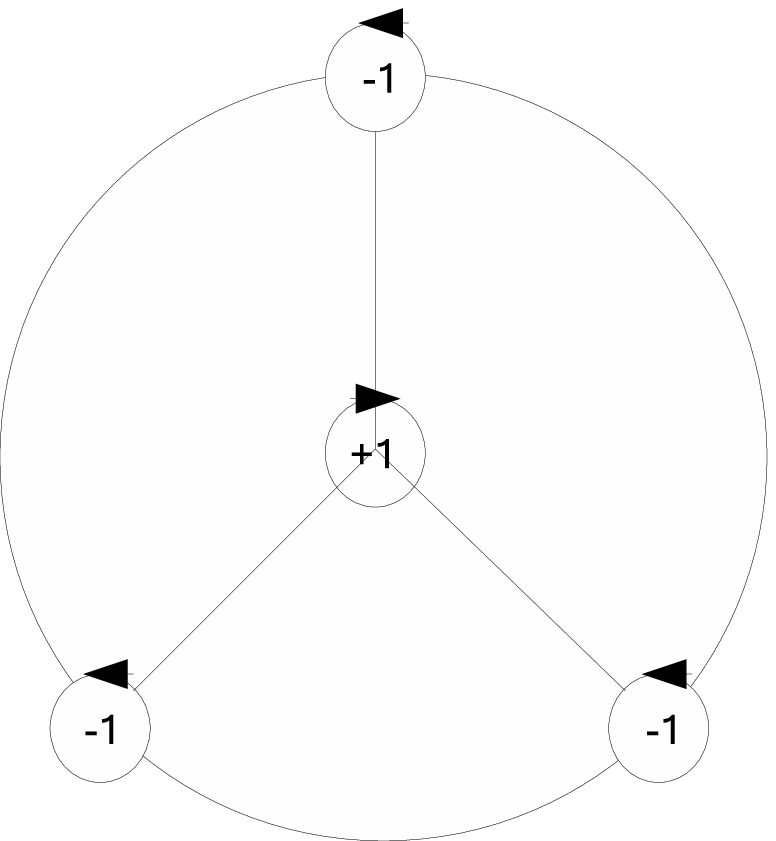}}
\caption{\label{fig.chargescheme} Sketch of the co-rotating topological charges observed in the azimuthon family of
Fig.~\ref{fig.triple}.}
\end{figure}

From a topological point of view, the above findings are somehow 
 surprising  because the two limiting vortices have
different charge ($q=1$ and $q=-2$). It is possible to understand this
 interesting feature when looking at the azimuthon
close to the respective vortices. Starting from the double charged vortex ($q=-2$), 
the azimuthon is created by adding a
counter-rotating single charged vortex ($m+q=1, $ see right panel in Fig.~\ref{fig.internalmodes3}).
However small the amplitude of this
single charged vortex might be, in the vicinity of the origin it will always be 
dominant as it grows as $\sim r$, whereas
the double charged vortex as $\sim r^2$. Thus, the azimuthon has a $q=1$ vorticity at the 
origin, and on a ring where
the amplitudes of the two vortices are equal lie three phase singularities with charge $-1$ (see
Fig.~\ref{fig.chargescheme} for a schematic sketch). As we can see in
Fig.~\ref{fig.triple}, the radius of this ring grows when we follow the family of azimuthons towards the single charged
vortex, and
the three singularities with total charge $-3$ move far away from the origin and finally disappear when we approach the
vortex with total charge $q=1$ (see discussion below). It is important to note that these three phase singularities have fixed positions with
respect to the position of the three humps and follow the amplitude rotation of the azimuthon (co-rotating).

We will now discuss in a greater detail 
 the three co-rotating phase singularities, 
 in particular, how they disappear when we approach
the vortex with total charge $q=1$. To this end, we analyze the asymptotic behavior 
for large $r$ of the three
components of the azimuthon $V$, $\delta V_{1}$ and $\delta V_{2}$ [see Eq.~(\ref{azimvor})].
For sufficiently large $r$, the convolution term $\theta$ in Eqs.~(\ref{NLS}) and (\ref{Eigen_vortex}) can be neglected when compared to
the term $\sim 1/r^2$ in the transverse Laplacian. Then, using the
modified Bessel functions, one can find the asymptotic behavior of
the involved functions easily:
\begin{subequations}
\begin{align}
\label{asymp}
\delta V_{1} & \sim
\frac{1}{\sqrt{r}}e^{-\sqrt{\kappa+\lambda_{0}}r}\left[1+\frac{
4\left(m+q\right)^{2}-1}{8r\sqrt{\kappa+\lambda_{0}}}+\mathcal{O}\left(\frac{1}{r^{2}}\right)\right] \\
\delta V_{2} & \sim
\frac{1}{\sqrt{r}}e^{-\sqrt{\lambda_{0}-\kappa}r}\left[1+\frac{
4\left(m-q\right)^{2}-1}{8r\sqrt{\kappa+\lambda_{0}}}+\mathcal{O}\left(\frac{1}{r^{2}}\right)\right] \\
V & \sim
\frac{1}{\sqrt{r}}e^{-\sqrt{\lambda_{0}}r}\left[1+\frac{4q^{2}-1}{8r\sqrt{\lambda_
{0}}}+\mathcal{O}\left(\frac{1}{r^{2}}\right)\right].
\end{align}
\end{subequations}
To find the radius $r_s$ where the phase singularities appear, one has to equal the amplitudes in the following manner:
\begin{equation}
\label{ringcondition}
\left|V\left(r\right)\right|=A_{r}\left|\delta V_{1}\left(r\right)+\delta
V_{2}\left(r\right)\right|.
\end{equation}
It is obvious from Eqs.~(\ref{asymp}) that such a radius exists 
for arbitrary small $A_r$, because one of the
$\delta V_i$ decays always slower than $V$ for $r\rightarrow\infty$. In our 
example we have $\kappa>0$, and therefore
$\delta V_2$ is responsible for creating our three co-rotating phase 
singularities. For $A_r\rightarrow0$ we find that
$r_s\rightarrow\infty$, the singularities move to infinite distances 
form the origin and (formally) vanish for $A_r=0$.
However, for practical observations in, e.g., numerical 
simulations those co-rotating phase singularities become
irrelevant when the surrounding amplitude becomes small.

\begin{figure}
\centerline{\includegraphics[width=.3\columnwidth]{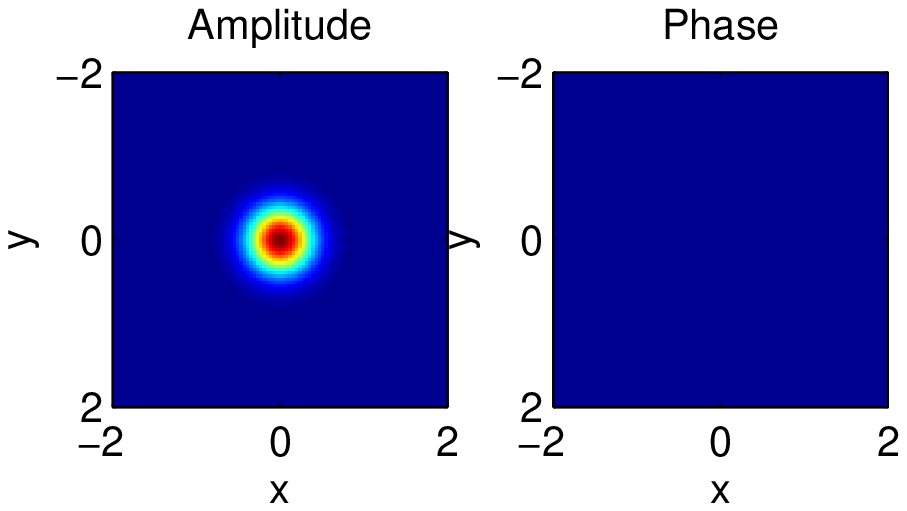}
\hspace{0.25cm}\includegraphics[width=.3\columnwidth]{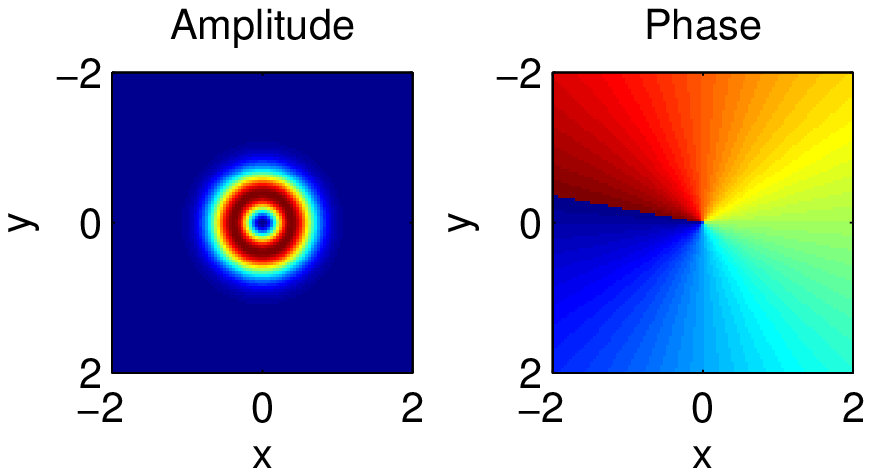}
\hspace{0.25cm}\includegraphics[width=.3\columnwidth]{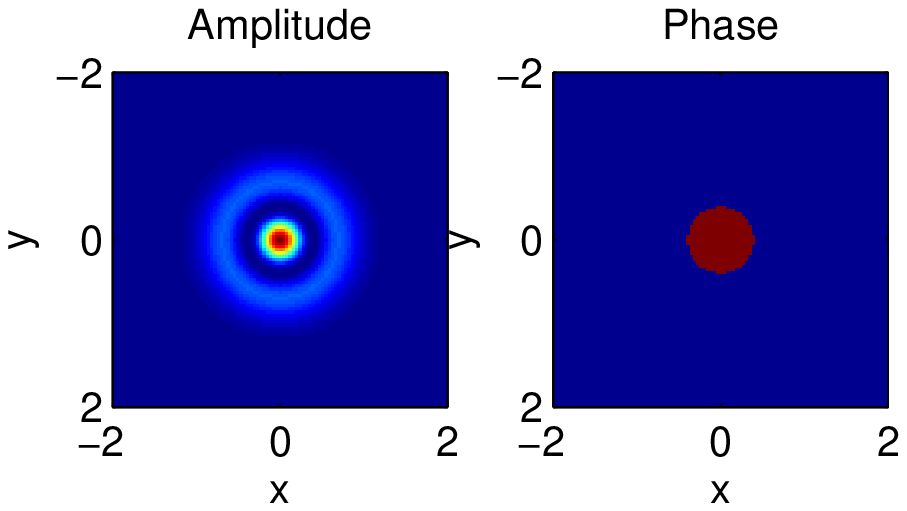}}
\vspace{0.25cm}
\centerline{\includegraphics[width=.3\columnwidth]{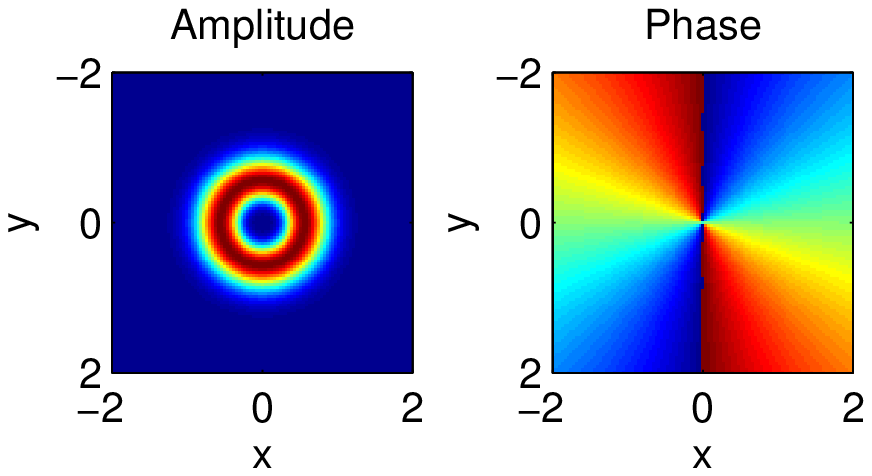}
\hspace{0.25cm}\includegraphics[width=.3\columnwidth]{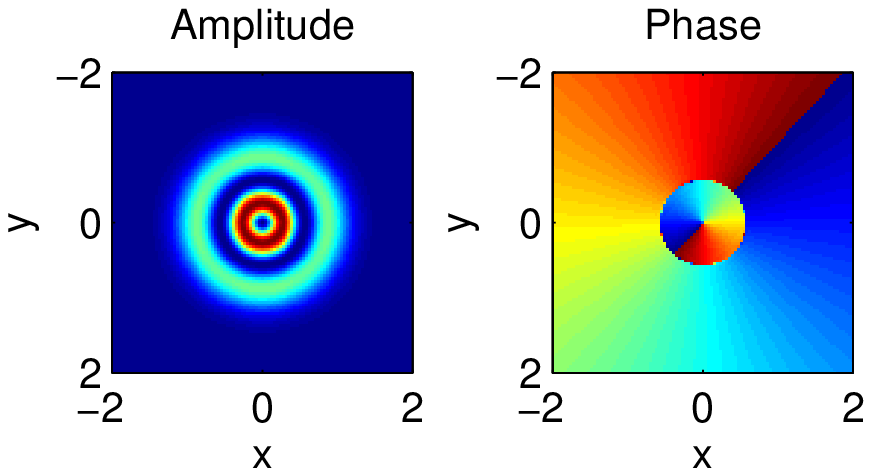}
\hspace{0.25cm}\includegraphics[width=.3\columnwidth]{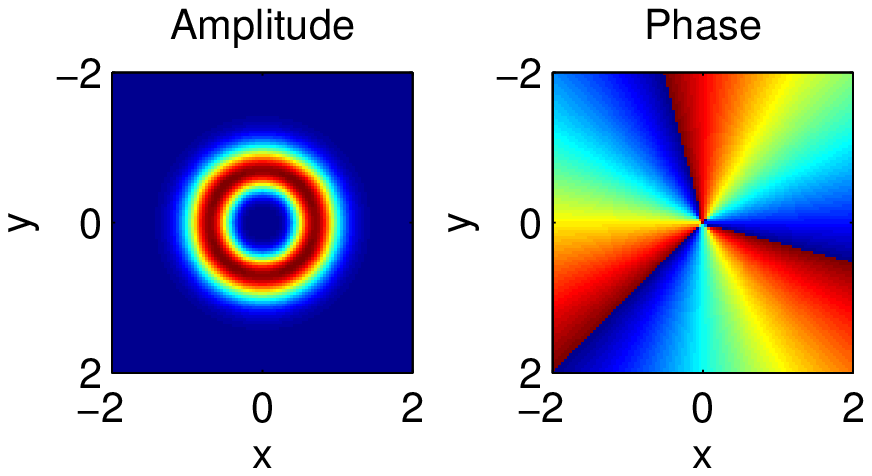}}
\caption{\label{fig.linearmodes} First six linear modes of a Gaussian potential with mass 630: Ground state $G_1(r)$,
$\lambda_{G_1}=86.6$; single charged vortex
$G_2(r)\exp(i\phi)$, $\lambda_{G_2}=73.5$; humped ring state  $G_3(r)$,
$\lambda_{G_3}=61.5$; double charged vortex
$G_4(r)\exp(i2\phi)$, $\lambda_{G_4}=60.9$; single charged double ring $G_5(r)\exp(i\phi)$, $\lambda_{G_5}=50$; triple
charged vortex $G_6(r)\exp(i3\phi)$, $\lambda_{G_6}=48.9$ (from left to right, top to down).}
\end{figure}

The observation that the above triple hump azimuthon has almost constant 
angular frequency when we follow the family
for constant mass $M$  regime can be explained going over to the highly nonlocal
limit. This rotation is not a purely nonlinear
phenomenon, but is mainly
a consequence  of mode beating. Let us have a look at the related 
linear limit where we replace the nonlocal response
$\Theta$ by the Gaussian kernel times mass $M$ (similar to Snyder-Mitchel model \citep{Snyder:sc:276:1538}). In this linear problem we can find
several eigenmodes (see Fig.~\ref{fig.linearmodes}). Mode beating between single ($q=1$) and
double ($q=-2$) charged vortices predicts $\Omega\sim(\lambda_{G_4}-\lambda_{G_2})/3=-4.2$, 
which is not too far from the
rotation frequency observed in the nonlocal nonlinear problem.

Another evidence that the linear contribution to 
the rotation of the triple hump dominates is that $|\Omega|$ increases strongly with
mass $M$ (and $\lambda$), as expected from $\Omega\sim(\lambda_{G_4}-\lambda_{G_2})/3$. E.g., for $M=200$ we find
$\Omega \sim -1.2$. In contrast to that, the double hump azimuthon
connecting single charged vortex and stationary dipole shows 
almost no dependency of $\Omega$ on the mass
\citep{Skupin:oe:16:9118}. This resembles the fact that in the linear problem mentioned above,
 mode beating predicts
$\Omega\sim(\lambda_{G_2}-\lambda_{G_2})/2=0$ for this structure. In fact,
 following a reasoning similar to
\cite{Buccoliero:ol:33:198}, we can identify in the 
expression (\ref{omega_exact}) for the rotation frequency a linear
and nonlinear contribution, $\Omega=\Omega_{ln}+\Omega_{nln}$.  
In the limit $\Theta=M\exp(-r^2/2)/2\pi$ and $U$ a
superposition of ''linear'' modes we readily see that $MN^{\prime}-L_zN=0$, and 
any rotation is due to
\begin{equation}
\label{omega_ln}
\Omega_{ln}=\frac{MI^{\prime}-L_zI}{L_z^2-MM^{\prime}}.
\end{equation}
In the special case that $U$ is a superposition of degenerated 
linear modes, we find $MI^{\prime}-L_zI=0$ and thus
$\Omega_{ln}=0$. However, if we consider the original nonlinear system where $\Theta$ is given by Eq.~(\ref{NL_R_int}),
an additional (nonzero) nonlinear contribution
\begin{equation}
\label{omega_nln}
\Omega_{nln}=\frac{MN^{\prime}-L_zN}{L_z^2-MM^{\prime}}
\end{equation}
to the rotation frequency occurs.

\begin{figure}
\centerline{\includegraphics[width=.3\columnwidth]{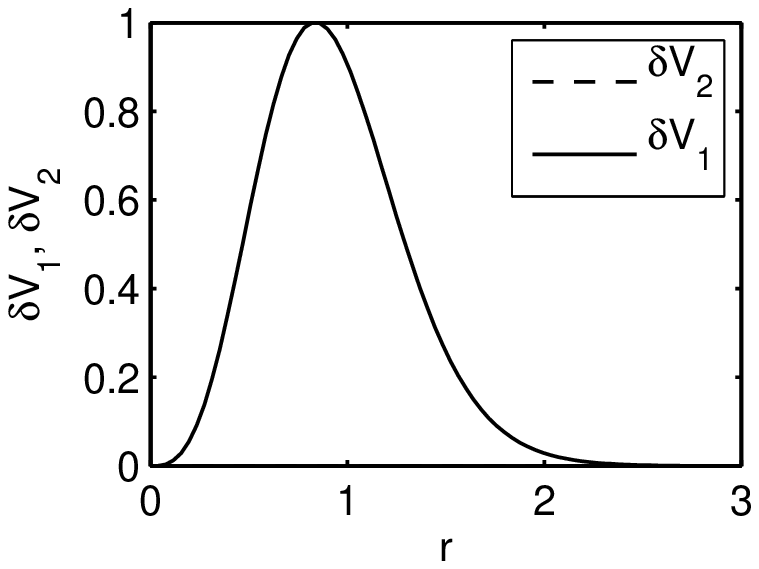}
\hspace{0.25cm}\includegraphics[width=.3\columnwidth]{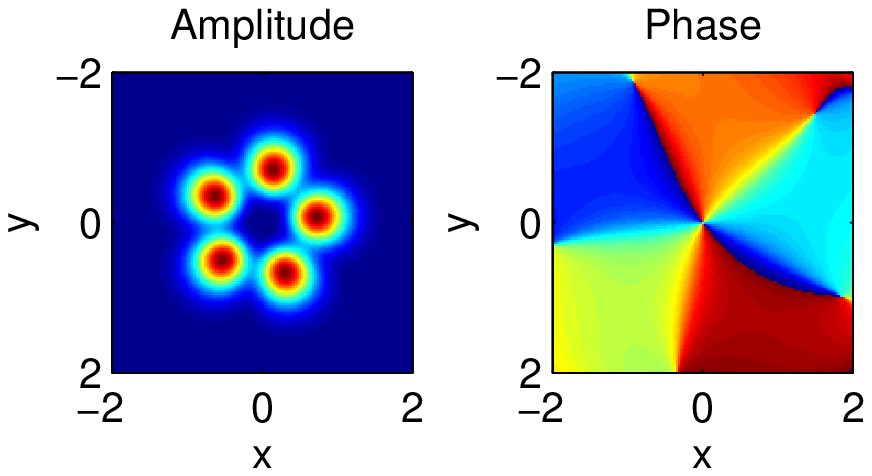}}
\caption{\label{fig.v2intdr5} Internal mode ($m=5$) of the double charged vortex ($q=-2$) with eigenvalue $\kappa=-8.6$,
and corresponding emanating azimuthon ($M=630$, $\alpha=0.56$).
$\delta V_{1}$ clearly dominates the small $\delta V_{2}$ component in the left picture.
}
\end{figure}

Once we have computed the internal mode of a vortex, we can construct all azimuthons branching from it. For example,
Fig.~\ref{fig.v2intdr5} shows a rotating five-hump azimuthon emanating
 from our double charged vortex ($q=-2$). The
corresponding internal mode shows a typical $~r^{(5-2)}$ dependence
 near the origin in $\delta V_1$, the amplitude of
$\delta V_2$ is very small. Hence, for the azimuthon, we see 
the double charged phase singularity ($q=-2$) of the vortex
in the origin. As observed for the rotatin triple-hump above,
 five singularities with $q=1$ appear on a ring, and they
are expected to move inwards when we follow the azimuthon family towards the triple charged vortex with $q=3$.

\begin{figure}
\centerline{\includegraphics[width=.3\columnwidth]{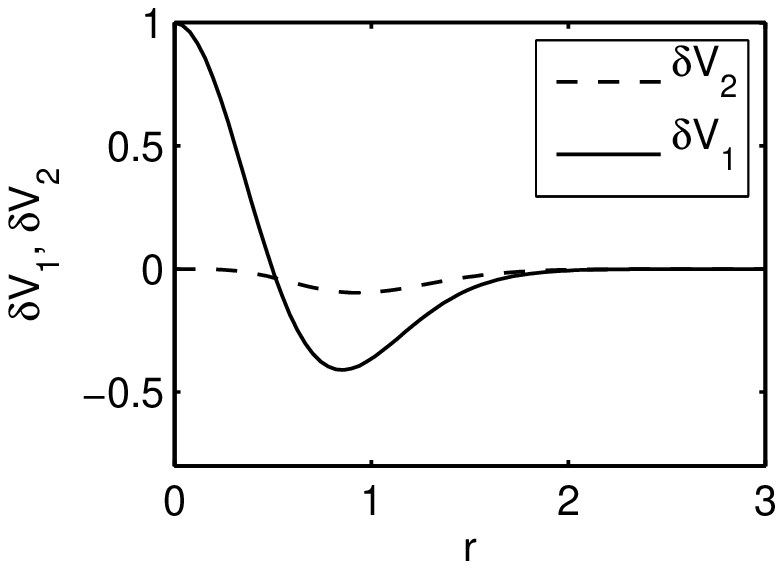}
\hspace{0.25cm}\includegraphics[width=.3\columnwidth]{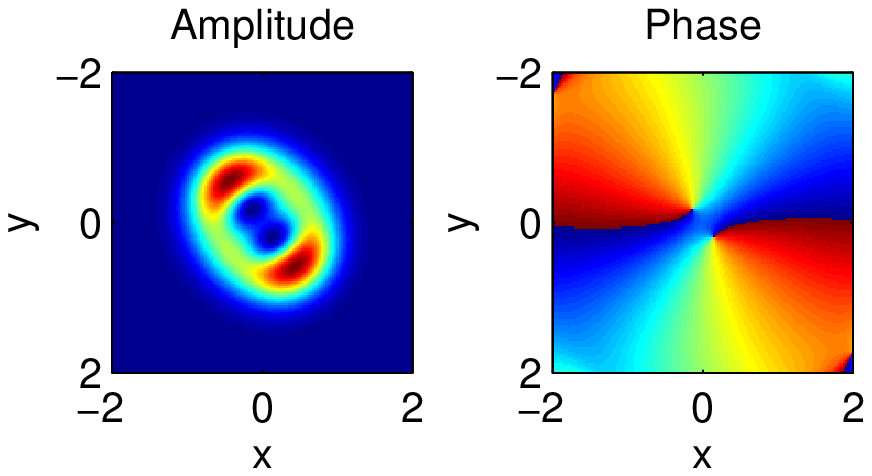}}
\caption{\label{fig.v2intdr} Internal mode ($m=2$) of the double charged vortex ($q=-2$) with eigenvalue
$\kappa=3.1$,
and corresponding emanating azimuthon ($M=630$, $\Omega=-1.44$).}
\end{figure}

We can also easily identify families of azimuthons previously found using a special ansatz. For instance, the double
charged vortex ($q=-2$) shows
another internal mode for $m=2$ with eigenvalue $\kappa=-3.1$. As can be seen in Fig.~\ref{fig.v2intdr}, the resulting
azimuthon belongs to the family connecting Hermite-Gaussian and Laguerre-Gaussian
self-trapped modes HN$_{20}$ and LN$_{20}$ \citep{Buccoliero:ol:33:198}. Note that for this solution our
definition of the structural parameter $\alpha$ does not make sense, hence in the caption of Fig.~\ref{fig.v2intdr} we
give the rotation frequency $\Omega$ instead, to characterize the azimuthon.

\begin{figure}
\centerline{\includegraphics[width=.3\columnwidth]{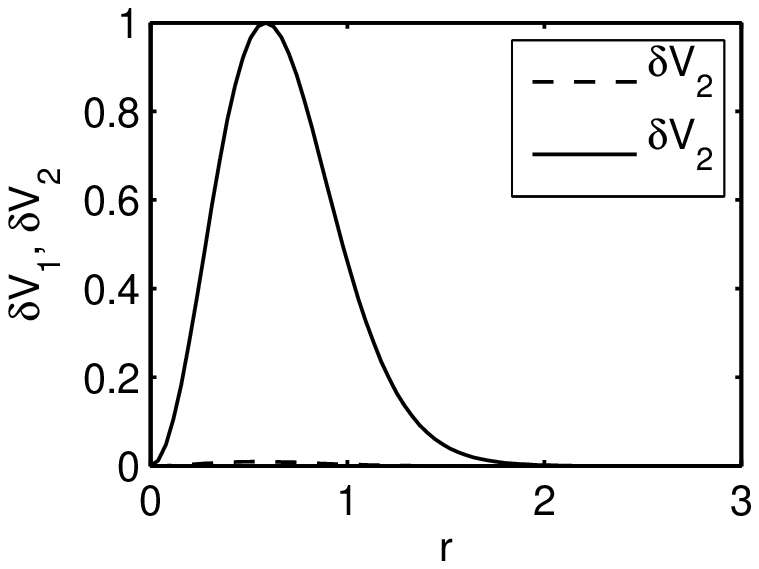}
\hspace{0.25cm}\includegraphics[width=.3\columnwidth]{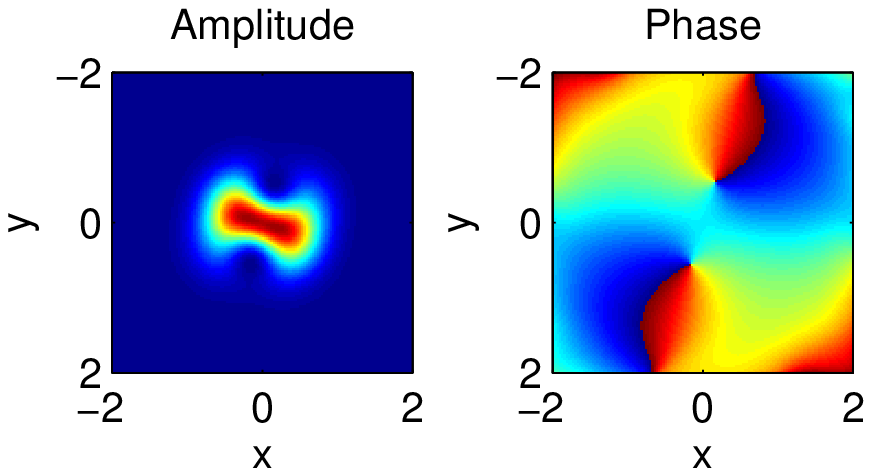}}
\caption{\label{fig.v2intdrbone} Internal mode ($m=2$) of the groundstate ($q=0$) with eigenvalue
$\kappa=-23.3$, and corresponding emanating azimuthon ($M=630$, $\Omega=14.3$).}
\end{figure}

Last but not least, we want to note here that the concept of azimuthons 
branching from solitons is not limited to
vortices. E.g., the single-hump ground state ($M=630$, $q=0$) features a $m=2$ internal mode with $\kappa=-23.3$. The
emanating azimuthon looks like a rotating bone, and possesses two 
phase singularities with opposite charges on an axis
perpendicular to the ''bone'' axis (see Fig.~\ref{fig.v2intdrbone}). The very high 
rotation frequency is again a
manifestation of linear mode beating, we find $\Omega\sim(\lambda_{G_1}-\lambda_{G_4})/2=12.9$.

If we reduce mass and therefore, in our scaling, reduce nonlocality these results may change. First of all, solitons
and azimuthons are expected to become unstable. Moreover, we observe that the
second component of the
perturbation becomes larger in amplitude when we leave the highly nonlocal regime (note that it is almost invisible in
Figs.~\ref{fig.internalmodes3}, \ref{fig.v2intdr5}, \ref{fig.v2intdr5}, and \ref{fig.v2intdrbone}). Also, certain types
of internal modes may vanish or new ones may appear. Hence, careful analysis of the internal spectra of soliton solutions is necessary to predict structure of azimuthons in a given regime or model, e.g., the local nonlinear Schr\"odinger equation.

\section{Conclusion}
\label{conclusion}
In conclusion, we have demonstrated a simple method for identifying rotating solutions in nonlocal nonlinear media.
We computed azimuthon solutions and their
rotation frequencies numerically and  showed that in the limit of minimal azimuthal amplitude
modulation, i.e., close to a vortex soliton, the rotation frequency is 
determined uniquely by eigenvalues of
the bound modes of the linearized version of the respective stationary nonlocal  solution. 
Moreover,  the intensity
profile  of the resulting azimuthons can be constructed from the corresponding linear eigensolution.  This offers  a
straightforward and  exhaustive method to identify
rotating soliton solutions in a given nonlinear medium.
At least, in the highly nonlocal regime, we find families of azimuthons which connect vortex solitons with different topological charge.

\begin{acknowledgements}
This research  was supported by the Australian Research Council.
 Numerical simulations were performed on the SGI Altix
3700 Bx2 cluster of the Australian Partnership for Advanced Computing (APAC).
\end{acknowledgements}

\bibliographystyle{spbasic}
\bibliography{references}

\end{document}